\documentclass[trackchanges]{emulateapj}

\usepackage{graphicx}
\usepackage{mathrsfs}	
\usepackage{amsmath}
\usepackage{color}
\usepackage{hyperref}
\usepackage{multirow}

\def\farcsec{\hbox{$.\!\!^{\prime\prime}$}}

\newcommand{\RNum}[1]{\uppercase\expandafter{\romannumeral #1\relax}}

\begin{document}

\title{Discovery of a very bright and intrinsically very luminous, strongly lensed Lyman alpha 
emitting galaxy at {\small z}=2.82 in the BOSS EMISSION-LINE LENS SURVEY$^{\dag}$}

\altaffiltext{$^{\dag}$}{Based on observations made with the Gran Telescopio Canarias (GTC) and William Herschel Telescope (WHT), in the Spanish Observatorio del Roque de los Muchachos of the IAC, under Director’s Discretionary Time
(DDT programs IDs: GTC2016-054 and DDT2016-077).}

\author{\mbox{Rui Marques-Chaves\altaffilmark{1,2}}}
\author{\mbox{Ismael P\'{e}rez-Fournon\altaffilmark{1,2}}}
\author{\mbox{Yiping Shu\altaffilmark{3}}}
\author{\mbox{Paloma I. Mart\'\i nez-Navajas\altaffilmark{1,2}}}
\author{\mbox{Adam S. Bolton\altaffilmark{4, 5}}}
\author{\mbox{Christopher S. Kochanek\altaffilmark{6}}}
\author{\mbox{Masamune Oguri\altaffilmark{7, 8, 9}}}
\author{\mbox{Zheng Zheng\altaffilmark{4}}}
\author{\mbox{Shude Mao\altaffilmark{10, 3, 11}}}
\author{\mbox{Antonio D. Montero-Dorta\altaffilmark{4}}}
\author{\mbox{Matthew A. Cornachione\altaffilmark{4}}}
\author{\mbox{Joel R. Brownstein\altaffilmark{4}}}

\altaffiltext{1}{Instituto de Astrof\'\i sica de Canarias, C/V\'\i a L\'actea, s/n, E-38205 San Crist\'obal de La Laguna, Tenerife, Spain}
\altaffiltext{2}{Universidad de La Laguna, Dpto. Astrof\'\i sica, E-38206 La Laguna, Tenerife, Spain}
\altaffiltext{3}{National Astronomical Observatories, Chinese Academy of Sciences, A20 Datun Rd., Chaoyang District, Beijing 100012, China}
\altaffiltext{4}{Department of Physics and Astronomy, University of Utah,
115 South 1400 East, Salt Lake City, UT 84112, USA}
\altaffiltext{5}{National Optical Astronomy Observatory, 950 N. Cherry Ave., Tucson, AZ 85719 USA}
\altaffiltext{6}{Department of Astronomy \& Center for Cosmology and Astroparticle Physics, Ohio State University, Columbus, OH 43210, USA}
\altaffiltext{7}{Research Center for the Early Universe, University of Tokyo, 7-3-1 Hongo, Bunkyo-ku, Tokyo 113-0033, Japan}
\altaffiltext{8}{Department of Physics, University of Tokyo, 7-3-1 Hongo, Bunkyo-ku, Tokyo 113-0033, Japan}
\altaffiltext{9}{Kavli Institute for the Physics and Mathematics of the Universe (Kavli IPMU, WPI), University of Tokyo, Chiba 277-8583, Japan}
\altaffiltext{10}{Physics Department and Tsinghua Centre for Astrophysics, Tsinghua University, Beijing 100084, China}
\altaffiltext{11}{Jodrell Bank Centre for Astrophysics, School of Physics and Astronomy, The University of Manchester, Oxford Road, Manchester M13 9PL, UK}

\setcounter{footnote}{5}

\begin{abstract}

We report the discovery of a very bright ($r = 20.16$), highly magnified, and yet intrinsically very luminous 
Ly$\alpha$ emitter (LAE) at $\rm z = 2.82$. This system comprises four images in the observer plane with a 
maximum separation of $\sim 6''$ and it is lensed by a $\rm z=0.55$ massive early-type galaxy. It was initially 
identified in the Baryon Oscillation Spectroscopic Survey (BOSS) Emission-Line Lens Survey for GALaxy-Ly$\alpha$ 
EmitteR sYstems (BELLS GALLERY) survey, and follow-up imaging and spectroscopic observations using the {\it Gran 
Telescopio Canarias} (GTC) and {\it William Herschel Telescope} (WHT) confirmed the lensing nature of this system.
A lens model using a singular isothermal ellipsoid in an external shear field reproduces quite well the main 
features of the system, yielding an Einstein radius of 2$\farcsec$95 $\pm$ 0$\farcsec$10, and 
a total magnification factor for the LAE of $8.8 \pm 0.4$. This LAE is one of the brightest and most luminous 
galaxy-galaxy strong lenses known. We present initial imaging and spectroscopy showing the basic physical and 
morphological properties of this 
lensed system.

\end{abstract}


\section{Introduction} \label{sec:intro}

The study of the physical properties of typical $L^{*}$ high-redshift galaxies has been limited by their faintness 
($r \simeq 24.5$ at $\rm z \sim 3$), even for 8-/10-m class telescopes. Over the past years the properties of these 
high-redshift galaxies have been studied by building large samples of hundreds to thousands individual spectra to 
construct high signal-to-noise (S/N) composite spectra \citep[e.g.][]{shapley}. Although these techniques have been 
very successful, they require a large amount of observing time and only probe the average physical properties of 
these galaxies.
Another way to study in detail high-redshift galaxies is using the fortuitous alignments with foreground massive 
structures which provide natural magnification and associated amplification produced by strong gravitational lensing. 
Hundreds of strongly lensed galaxies have been discovered in the last years employing various observational techniques, 
mainly from optical to radio. However, only a handful of optically very bright ($r \sim 20$), strongly magnified 
high-redshift galaxies, have been discovered so far \citep{yee, allam, smail, belo, lin, dahle} allowing detailed 
spectroscopic studies of their individual properties, such as stellar populations, chemical composition, and kinematics 
of the interstellar medium \citep[e.g.,][]{pettini2000, pettini2002, quider2009, quider2010, des}. 

In this Letter, we report the discovery of a bright ($r \sim 20$) quadruply gravitationally lensed Ly$\alpha$ 
emitter (LAE). We provide the first physical and morphological analysis of the lensing galaxy and the lensed LAE. 
Throughout the Letter, we adopt a cosmology  with $\Omega_{\rm m}=0.274$, $\Omega_\Lambda=0.726$ and $H_o=70$\,km\,s$^{-1}$\,Mpc$^{-1}$. All quoted 
magnitudes are in the AB system.

\section{Discovery and follow-up} \label{sec:disc}

\begin{figure*}[ht]
\figurenum{1}
\includegraphics[width=180mm,scale=1]{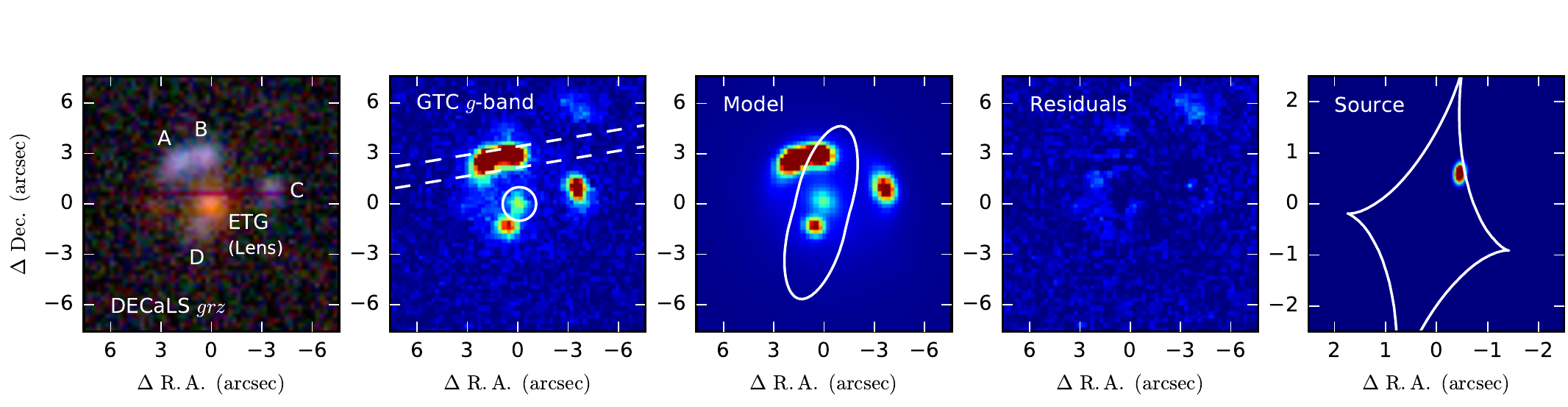}
\caption{From left to right: DECaLS $grz$ color image of the lens system; GTC/OSIRIS $g$-band 
image showing the orientation of the WHT/ACAM and GTC/OSIRIS long slits (white dashed lines) and the position of the spectroscopic $1''$-radius BOSS fibre (white circle); predicted lensed and foreground images with the critical line; 
final residuals from the best-fit model; and position of the LAE in the source plane relative to the caustic. All 
the images are centered on the lensing galaxy and oriented such that North is up and East is to the left.\label{fig:image}}
\end{figure*}

In the past years, by searching for secondary emission lines in the spectra of massive elliptical galaxies within 
the database of the Sloan Digital Sky Survey \citep[SDSS:][]{york}, the Sloan Lens ACS Survey \citep[SLACS: e.g.,][]{bolton2006, bolton2008}, and the Baryon Oscillation Spectroscopic Survey (BOSS) Emission-Line Lens Survey \citep[BELLS:][]{Brownstein12} have discovered over 100 lensed star-forming galaxies at moderate redshifts ($\rm z 
\sim 1$). 
Very recently, by applying spectroscopic selection techniques similar to SLACS and BELLS, but adapted to 
high-redshift Ly$\alpha$ emission, \cite{shu_a} identified a new sample of 187 high-probability lensed LAE candidate 
systems at 2 $<$ $\rm z_{LAE}$ $<$ 3, the BOSS Emission-Line Lens Survey for GALaxy-Ly$\alpha$ 
EmitteR sYstems (BELLS GALLERY) survey, selected from the final data release (DR12) of the BOSS \citep{Dawson13} of the SDSS-III \citep{SDSSIII}.
Of these, 21 highest-quality candidates were recently observed with the {\it Hubble Space Telescope} (HST) and the 
first results were presented in \cite{shu_b}.

By visual inspection of SDSS and DECaLS\footnote{Dark Energy Camera Legacy Survey: 
\url{http://legacysurvey.org/decamls/}} color images of the BELLS GALLERY sample, we found one candidate showing 
bluish features $\simeq$ 3$\farcsec$3 from the $\rm z=0.5531$ massive early-type galaxy, SDSS J142954.80+120235.6 
(hereafter ETG). Its BOSS spectrum (Plate-MJD-Fiber: 5463-56003-121) shows a secondary emission line at $4652 \rm 
\AA$ \citep[Ly$\alpha$ at $\rm z  = 2.8253$;][]{shu_a}. The positions of the bluish features with respect to the ETG 
are consistent with a $\it{fold}$ lensing configuration: a bright lensed image pair, A and B (with a separation of 
$\simeq $1$\farcsec$5), and two fainter images, C and D (see Fig. \ref{fig:image}). The lensed image pair, A and B, 
is identified in SDSS as a single source, SDSS J142954.88+120238.3 (hereafter BG1429+1202 for the lensed LAE, where 
BG stands for BELLS GALLERY), showing blue colors in DECaLS and SDSS bands (Table \ref{tab1}). The lensed 
A and B images are also detected in the UKIRT Infrared Deep Sky Survey \citep[UKIDSS:][]{lawrence} Large Area Survey 
(LAS), only in Y-band with $21.06 \pm 0.19$ mag and $21.12 \pm 0.20$ mag, respectively.

We carried out Director Discretionary Time (DDT) for optical imaging and long-slit spectroscopic observations on 2016 
June 29 using the Auxiliary-port Camera \citep[ACAM:][]{acam} at the \textit{William Herschel Telescope} (WHT) 
to confirm the lensing nature of this LAE. 
The ACAM long-slit was oriented at sky position angle (PA) = $103^{\circ}.35 $, and positioned so as to encompass the 
two brightest lensed images, A and B, as shown in Fig. \ref{fig:image}. Despite the bad seeing conditions that night 
($\sim 4''$ FWHM), we could confirm the lensing nature of this system with the detection in the long-slit spectrum of 
strong rest-frame UV continuum (with Si {\sc iv} and C {\sc iv} in absorption) and Ly$\alpha$ emission at a redshift 
$\rm z = 2.823 \pm 0.008$, in agreement with the redshift of the Ly$\alpha$ detected in the BOSS fiber spectrum, 
likely from the lensed image D (see Fig. \ref{fig:image}).

\begin{table}
\begin{center}
\caption{Properties of the system\label{tab1}}
\begin{tabular}{c c c c c c}
\hline \hline
ID & R.A.$^{\rm (1)}$  & Dec.$^{\rm (1)}$  & $g^{\rm (2)}$ & $r^{\rm (3)}$ & $z^{\rm (4)}$  \\ 
 & (J2000.0)  &  (J2000.0) &  &  &  \\
\hline
ETG		& 14:29:54.806 & +12:02:35.53 & 22.88 & 20.51 & 19.00 \\
        &             &             & $\pm 0.12$ & $\pm 0.07$ & $\pm 0.10$ \\
A$^{\rm (5)}$   	& 14:29:54.936 & +12:02:38.23 & 21.48 & 21.23 & 21.32\\
        &             &             & $\pm 0.08$ & $\pm 0.03$ & $\pm 0.04$ \\
B$^{\rm (5)}$ 		& 14:29:54.831 & +12:02:38.48 & 21.64 & 21.34 & 21.31\\
        &             &             & $\pm 0.08$ & $\pm 0.03$ & $\pm 0.04$ \\
C       & 14:29:54.570 & +12:02:36.69 & 22.27 & 22.22 & 22.02\\
        &             &             & $\pm 0.09$ & $\pm 0.08$ & $\pm 0.09$ \\
D       & 14:29:54.848 & +12:02:34.24 & 22.57 & 22.30 & 21.85\\
        &             &             & $\pm 0.11$ & $\pm 0.10$ & $\pm 0.11$ \\
\hline 
Total LAE  &             &        & 20.40 & 20.16 & 20.07\\
\hline 
\end{tabular}
\\
\end{center}
\textsc{      Note.} --- $\rm (1)$ positions with 0$\farcsec$11 astrometric r.m.s. using GAIA DR1 \citep{gaia};
$\rm (2)$ photometry and 1$\rm \sigma$ errors from the GTC/OSIRIS $g$-band image; $\rm (3)$ and $\rm (4)$ photometry 
and 1$\rm \sigma$ errors from the DECaLS DR2 Tractor catalog, except for the lens which were taken from the SDSS DR12 
due to image artifacts in DECaLS co-adds, and $\rm (5)$ the lensed images A and B were deblended using {\sc GALFIT} 
\citep{galfit}. The magnitudes listed have not been corrected for the Galactic dust extinction, which is 0.104, 0.072, 
and 0.040 mag for $g$, $r$, and $z$, respectively \citep{schlafly2011}.
 \\
\end{table}

We observed again BG1429+1202 on 2016 July 29, this time in very good seeing conditions ($\simeq $0$\farcsec$75 
FWHM, measured in the 330 s $g$-band acquisition image), using the Optical System for Imaging and 
low-Intermediate-Resolution Integrated Spectroscopy camera (OSIRIS\footnote{\url{http://www.gtc.iac.es/instruments/osiris/}}) 
on the \textit{Gran Telescopio Canarias} (GTC). We used the R1000B grism, which provides a spectral coverage of 3630 - 
7500 $\rm \AA$ (950 - 1960 $\rm \AA$, rest-frame) and a dispersion of 2.12 $\rm \AA$ px$^{-1}$. The OSIRIS 1$\farcsec$2 
wide long-slit was oriented in the same PA used in the WHT/ACAM spectroscopic observations (Fig. \ref{fig:image}). 
Given this configuration, the corresponding spectral resolution is $\simeq 8 \rm \AA$ (or $\simeq 500$ km s$^{-1}$ 
FWHM). The total integration time was 44 minutes, split into 4 $\times$ 660 s. The data were reduced with {\sc iraf} 
and {\sc python} tasks.

\begin{figure*}
\figurenum{2}
\centering
\includegraphics[width=160mm,scale=1]{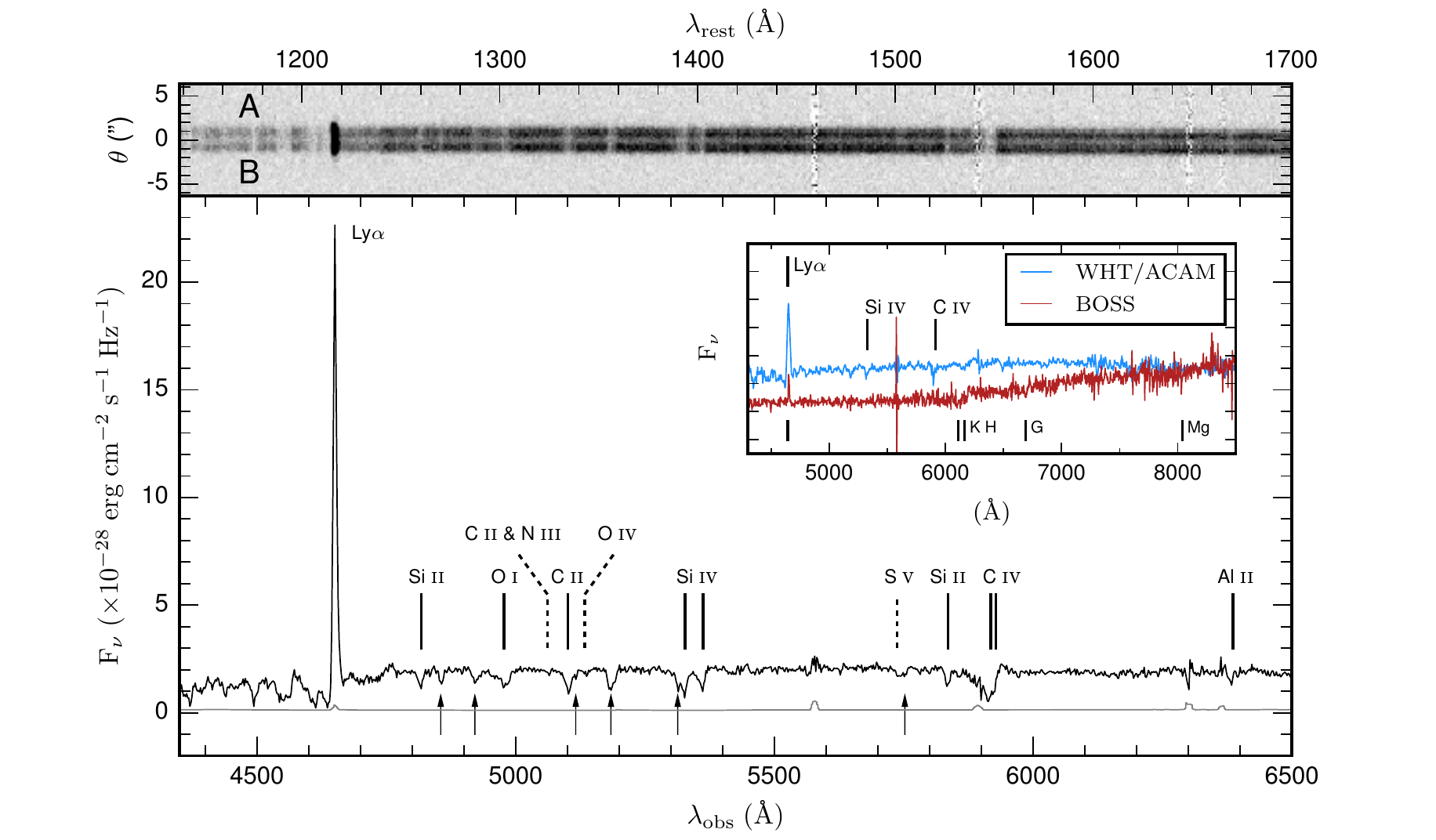}
\caption{GTC/OSIRIS spectrum of BG1429+1202 (black solid line) and its $1 \sigma$ error (grey solid line). Ticks 
mark the positions of the strong interstellar absorption features (solid lines) and photospheric lines (dashed lines) 
of BG1429+1202, as well as other absorption features related to intervening systems at lower redshifts (black solid 
arrows). The 2D spectrum is also shown on the top. In the inset panel we show the BOSS and WHT spectra with the positions 
of the Ly$\alpha$ emission as well as the absorption features related to the ETG at $\rm z=0.5531$ from the BOSS fiber 
spectrum.\label{fig:spec}}
\end{figure*}

\section{Analysis and Discussion} \label{sec: disc}

\subsection{Source Properties} \label{sec: source}

The rest-frame UV spectrum of the brightest lensed images A and B, is shown in Fig. \ref{fig:spec}. Despite the 
relatively short exposure time, the achieved high S/N GTC/OSIRIS spectrum ($\sim 25$ in the continuum) shows a strong 
Ly$\alpha$ emission, and a series of strong absorption lines, similar to those seen in the composite spectrum of 
hundreds $\rm z \sim 3$ Lyman break galaxies (LBGs) \citep{shapley}. The strongest absorption features are associated 
to the interstellar medium and stellar winds in a variety of ionization states, from neutral (O {\sc i}) to highly 
ionized species (Si {\sc iv} or C {\sc iv}). High ionization features with a strong P Cygni profile are seen both in 
C {\sc iv} $\rm \lambda 1548$, 1550 and Al {\sc iii} $\rm \lambda 1854$, 1862 doublets, which is indicative of stellar 
winds from very young massive stars. However, we identify additional absorption features unrelated to BG1429+1202 
(nine absorption features associated to an intervening metal-line system at $\rm z_{\rm abs} =$ 2.179 
$\pm$ 0.001, and one broad absorption line at $5183 \rm \AA$ with an observed equivalent width $W_{\rm obs} = 
6.95 \pm 0.2$ $\rm \AA$ remains unidentified). 
Analysis of the spectra for the individual lensed images A and B shows no differences 
in the profiles of the absorption features neither evidence for velocity offsets between them, as expected if they 
are both images of the same background source. The same happens to their rest-frame UV slope, $\beta$, which is 
essentially flat in $F_{\nu}$. Adopting a simple power-law approximation for the UV spectral range $F_{\rm \lambda} 
\propto \rm \lambda^{\beta}$ and using the observed $r$ and $z$ magnitudes ($\sim 1600$ - $2400 \rm \AA$ rest-frame), 
we derive $\beta = -2.1 \pm 0.1$, which implies an effective UV extinction $A_{1600} \sim 0.6$ or $E(B-V) \sim 0.13$, 
reflecting modest dust content \citep{calzetti2000}.

We identified several photospheric absorption features, from which we derived the systemic redshift $\rm z_{\rm sys} = $
2.8224 $\pm$ 0.0013 using the cleanest among them: O {\sc iv} $\rm \lambda 1343$, and a close blend of 
C {\sc ii} and N {\sc iii} at $\rm \lambda 1324$. 

The Ly$\alpha$ emission line has an observed (A + B) flux of $F_{\rm Ly \alpha} = (2.1 \pm 0.3) \times 10^{-15}$ erg 
s$^{-1}$ cm$^{-2}$, much higher than the Ly$\alpha$ flux measured within the $1''$-radius BOSS fibre ($F_{\rm Ly 
\alpha}^{\rm BOSS} = 0.256 \times 10^{-15}$ erg s$^{-1}$ cm$^{-2}$). The measured rest-frame equivalent width of 
Ly$\alpha$ is $W_{0}^{\rm Ly \alpha} = 39 \pm 15$ $\rm \AA$. The errors reflect the uncertainty in the determination 
of the stellar continuum redward of Ly$\alpha$. Although the line appears unresolved in our low resolution spectrum 
($\rm FWHM \simeq 500$ km s$^{-1}$), it is resolved in the BOSS spectrum. Fitting a Gaussian to the BOSS Ly$\alpha$ 
line, we measured a FWHM of $382 \pm 50$ km s$^{-1}$, after accounting for the instrumental broadening. The nebular 
C {\sc iii]} $\rm \lambda$1906, 1908 doublet is also detected in emission but with low significance ($4 \rm \sigma$) 
and not resolved in our GTC/OSIRIS spectrum. 
This galaxy can be classified as a LAE, given the rest-frame equivalent width and the velocity width of the 
Ly$\alpha$ line \citep[e.g.,][]{ouchi2008}. From our spectrum and available photometric data we do not detect evidence 
of any AGN contribution. However, a more extensive characterization awaits higher spatial resolution imaging and 
multi-wavelength photometry.

\begin{figure*}
\figurenum{3}
\centering
\includegraphics[width=150mm,scale=1]{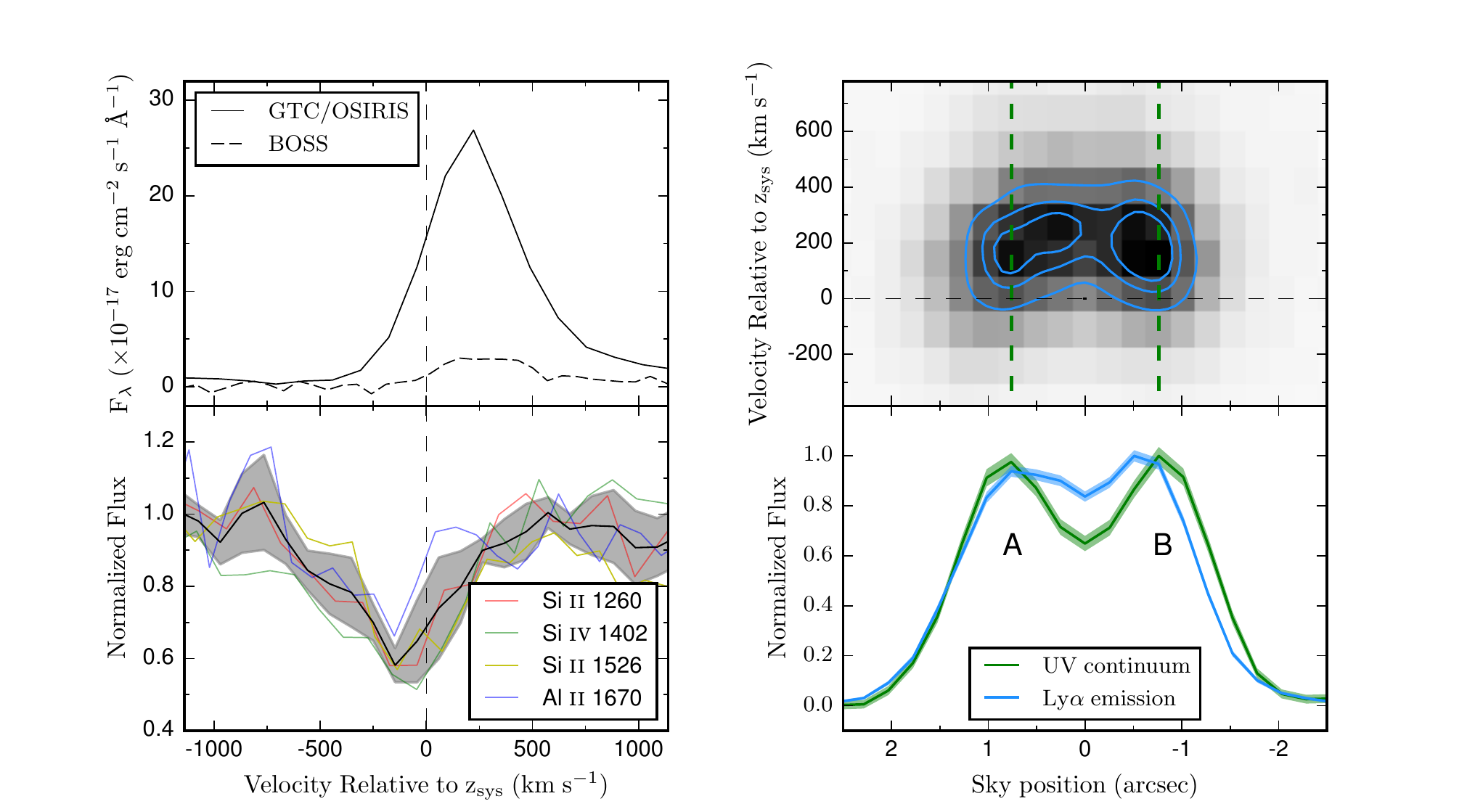}
\caption{Left: GTC/OSIRIS and BOSS Ly$\alpha$ emission (upper panel) and interstellar absorption features (lower panel) 
on rest velocity frame. The Ly$\alpha$ emission appears to be redshifted, $v_{\rm Ly \alpha} \sim +200$ km s$^{-1}$, 
while the interstellar absorption lines are blueshifted by $v_{\rm abs} \sim -150$ km s$^{-1}$ (mean profile in black 
with the standard deviation in grey shadow). Right: 2D Ly$\alpha$ emission in the 
position-velocity plane (upper panel) where a spatial gradient of velocity is seen in both A and B images. 
In the lower panel we show the normalized spatial profiles of the rest-frame UV continuum 
(green line) and Ly$\alpha$ emission (blue line). The shaded regions show the $1 \sigma$ errors of the spatial profiles. 
The Ly$\alpha$ emission appears more extended between A and B, than the rest-frame UV continuum.
\label{fig:vel}}
\end{figure*}

A careful analysis of the kinematics goes beyond the present Letter given the low spectral resolution of these data. 
However, the high S/N of the GTC/OSIRIS spectrum allows us to detect differences in the kinematics of the Ly$\alpha$ 
emission and interstellar features. Fig. \ref{fig:vel} (left panel) shows velocity plots of the Ly$\alpha$ emission 
line and several normalized interstellar absorption lines, relative to $\rm z_{\rm sys}$. The Ly$\alpha$ emission has 
its peak redshifted with a velocity $v_{\rm Ly \alpha} \sim +200$ km s$^{-1}$, while the interstellar absorption lines 
appear blueshifted by $\sim -150$ km s$^{-1}$, which is consistent with galaxy-scale outflows of material from the 
galaxy in the form a wind, similar to those seen in other star-forming galaxies at $z \sim 3$ \citep{shapley}.

We also notice that a spatial gradient of the Ly$\alpha$ velocity is present in the 2D spectrum (Fig. \ref{fig:vel}, 
right upper panel). This velocity structure is seen both in A and B but appears mirrored, as expected for those images 
in this system as A and B images straddle the fold critical curve (see figure \ref{fig:image}). We checked carefully 
the 2D spectrum to see if this pattern is present in other lines, but we found none with this signature. 
The 1D spatial distribution of Ly$\alpha$ is also compared with that of the rest-frame UV continuum (Fig. 
\ref{fig:vel}, right lower panel). The Ly$\alpha$ emission appears more extended in the inner region between A and B, 
than the UV continuum. Deeper and higher spatial resolution observations are needed to constrain the Ly$\alpha$ and UV continuum spatial distributions.

\subsection{Lens Model} \label{sec: lens}

To interpret the properties of the lensed system in more detail, we used the 330 s GTC/OSIRIS $g$-band image obtained 
in $\simeq 0.75''$ FWHM seeing for accurate lens modeling. As shown in Fig. \ref{fig:image}, the lens system comprises 
four images forming a so-called $\it{fold}$ configuration, when the source lies very close to a $\it{fold}$ caustic. 
Similar to previous works \citep{bolton2008, Brownstein12, shu15, shu_b, shu_c}, we have developed a lens model using 
a non-linear optimizer, consisting on minimizing a $\chi^{2}$ function using the Levenberg-Marquardt algorithm with 
the {\tt LMFIT} package \citep{lmfit}, where the observational data is compared to the model \citep[see][for details]{shu_b}.
The foreground-light is modeled using the elliptical S{\'e}rsic profile and similar to \cite{shu15, shu_b, shu_c} its 
subtraction is performed jointly with the lens modeling. The lens model includes a mass distribution of the foreground 
lens parameterized as a singular isothermal ellipsoid (SIE), and an additional external shear is included to model the 
higher-order effect from the environment. The surface brightness distribution of the background source is reconstructed 
parametrically using an elliptical S{\'e}rsic model. The foreground-light model is combined with the predicted lensed 
images and convolved with the point-spread function (PSF) which was modeled with a star in the GTC/OSIRIS $g$-band 
field-of-view.

The best fit lens model ($\chi^{2} \rm / dof = \rm 3494/3709$) predicts for the lens an Einstein radius $b_{\rm SIE} = 
$2$\farcsec$95$ \pm $0$\farcsec$10, minor-to-major axis ratio $q = \rm 0.34$ and a position angle 
$\rm P.A. = \rm 165.5$ deg. 
The strength and position angle of the external shear is $\gamma = 0.059 \pm 0.008$ and $\phi_{\gamma} = \rm 89.9$ 
deg, respectively. The characteristic lensing velocity dispersion defined as $\sigma_{\rm SIE} = c \sqrt{\frac{b_{\rm 
SIE}}{4 \pi} \frac{D_{\rm L}}{D_{\rm LS}}}$ is $390 \pm $6 km s$^{-1}$, where $D_{\rm LS}$ and $D_{\rm S}$ 
are the angular diameter distances from the lens and the observer to the source, respectively. We also find a 
minor-to-major axis ratio of the SIE component (0.34), smaller than that of the light distribution suggested by the 
$g$-band model result (0.85). 
The lensing velocity dispersion suggests that cluster or line-of-sight structures also contribute a substantial fraction 
of convergence. There is no clear evidence of a crowded environment around the ETG, either by visual inspection of 
color images or in the SDSS photometric redshifts, but $\sim 1'$ to the North there is a galaxy, SDSS J142953.71+120333.9, 
with a BOSS spectroscopic redshift $\rm z=0.5527 \pm 0.0002$, very close to the lensing ETG, indicating that a cluster or 
group of galaxies at $\rm z \simeq 0.55$ may be present, as suggested by the external shear field. 

For the source, the lens model gives local magnifications of 3.2, 3.1, 1.8 and 0.7 for images A, B, C and D, respectively, 
which means that the total magnification is $8.8 \pm 0.4$. The source has an effective radius $R_{\rm eff} = $ 0$\farcsec$159 
$\pm$ $\rm $0$\farcsec$007, which corresponds to $R_{\rm eff} = \rm 1.28 \pm \rm 0.06$ kpc for the adopted cosmology. 
The source has a minor-to-major axis ratio $q = 0.56$ and S{\'e}rsic index $\rm n = 3.9$. It is centered at $\Delta 
\rm R.A. = $-0$\farcsec$43 and $\Delta \rm Dec. = $0$\farcsec$56 relative to the center of the lens galaxy. 

\subsection{Intrinsic Properties} \label{sec: int}

Having determined the magnification of the LAE we can estimate its intrinsic properties. From the total DECaLS DR2 
$r$-band magnitude, we determine a rest-frame 1600 $\rm \AA$ luminosity $L_{1600} 
= (6.12 \pm 0.48) \times 10^{30}$ erg s$^{-1}$ Hz$^{-1}$. Using the \cite{kennicutt} conversion, this rest-frame UV 
luminosity translates into an intrinsic star formation rate (SFR) of $\simeq 90$ $M_{\odot}$yr$^{-1}$, when corrected for 
magnification, reddening and the lower proportion of low-mass stars in the \cite{chabrier} stellar 
IMF relative to the standard \cite{salpeter} adopted by Kennicutt (a factor of $1/1.8$).
Turning to Ly$\alpha$, assuming the space distribution of Ly$\alpha$ in the source 
plane and its lensing magnification are similar to that of the rest-frame UV continuum, and applying the correction for 
the magnification and the slit losses (the GTC/OSIRIS slit captured a fraction $\simeq 0.60$ of the total light of 
BG1429+1202), we derive an intrinsic Ly$\alpha$ luminosity $L_{\rm Ly \alpha} = (2.80 \pm 0.39) \times 10^{43}$ erg 
s$^{-1}$. 
Assuming case-B recombination and the \cite{kennicutt} conversion, this luminosity 
translates in a $\rm SFR(Ly\alpha) \simeq 25$ $M_{\odot}$yr$^{-1}$. Comparing the estimates of SFR from the rest-frame UV and 
Ly$\alpha$, we measure $f^{\rm Ly\alpha}_{\rm esc} \sim 0.30$, consistent with that estimated from LAEs at 
$\rm z \sim 3$ \citep[e.g.,][]{zheng2016, verhamme}. However, we should note that the measured 
$f^{\rm Ly\alpha}_{\rm esc}$ results from the assumption that the spatial distribution of Ly$\alpha$ (and its 
magnification) follows the rest-frame UV continuum. 
Ly$\alpha$ halos are hard to be resolved from individual LAEs, and have been studied mainly by using stacking techniques 
\citep[][]{steidel, momose}, or in nearby high-redshift analogs \citep[e.g.][]{yang}. However, for a few cases, strong gravitational lensing allows spatially resolved studies of high-redshift galaxies \citep[e.g.,][]{patricio}. 
A more detailed analysis will be possible with high-resolution narrow-band imaging and integral field spectroscopy.

\begin{table*}
\begin{center}
\caption{Intrinsic properties of BG1429+1202 and other bright galaxy-galaxy lenses \label{tab2}}
\begin{tabular}{c c c c c c c }
\hline \hline
Object & z  & $m_{\rm UV}^{\rm a}$  & $\mu^{\rm b}$ & $L_{\rm UV}^{\rm c}$ & $L_{\rm Ly \alpha}^{\rm d}$  & Reference \\ 
       &    &  ($\simeq \rm 1500 - 1700 \AA$) & & ($\times 10^{29}$ erg s$^{-1}$ Hz$^{-1}$) & ($\times 10^{42}$ erg s$^{-1}$) & \\
\hline
BG1429+1202 & 2.822 & 20.16 & 8.8 & 6.99 & 28.0 & -- \\
MS 1512-cB58 & 2.726 & 20.64 & 30  & 1.16 & -- & 1,2 \\
Cosmic Eye    & 3.073 & 20.30 & 28  & 2.08 & -- & 3\\
8 o'clock    & 2.735 & 19.22 & 12.3 & 10.55 & -- & 4 \\
Cosmic Horseshoe & 2.381 & 19.70 & 24 & 2.74  & 3.3 & 5,6 \\
LBGs         & $\sim 3$ &  24.61 & -- & 1.06  & -- & 7 \\
LAEs         & $\sim 3.1$ &  25.84 & -- & 0.36  & 5.8 & 8\\
\hline 
\end{tabular}
\\
\end{center}
\textsc{      Note.} --- $^{\rm (a)}$ rest-frame UV apparent magnitudes from $r$- or $i$-band, depending on the redshift; 
$^{\rm (b)}$ total magnification factor; $^{\rm (c)}$ and $^{\rm (d)}$ intrinsic rest-frame UV and $\rm Ly \alpha$ luminosity, 
respectively, corrected from the lensing magnification; 
 \\
\textsc{      References.} --- (1) \cite{ellingson}; (2) \cite{seitz}; (3) \cite{smail}; (4) \cite{allam}; (5) \cite{belo}; (6) \cite{quider2009};
(7) \cite{reddy2009}; (8) \cite{ouchi2008}.
 \\
\end{table*}

In order to establish how typical are the intrinsic properties of this galaxy, we compare it with other UV-selected 
$\rm z \sim 3$ LBGs and LAEs. BG1429+1202 is intrinsically more luminous in the rest-frame UV by a factor of $7$ and 
$19$, relatively to $L^{*}$ from the luminosity functions of \cite{reddy2009} for $\rm z \sim 3$ LBGs and $\rm z \sim 
3.1$ LAEs selected by narrow-band imaging by \cite{ouchi2008}, respectively. 
It is also intrinsically very luminous in $\rm Ly \alpha$ emission, when compared with 
$L_{\rm Ly \alpha}^{*}$ from the luminosity functions of LAEs at $\rm z \sim 3.1$ \citep[factor of $\sim 5$;][]{ouchi2008}
and LAEs at $\rm z = 2.8$ in CDFS \citep[factor of $\sim 9-5$;][]{zheng2016}.

A comparison is presented in table \ref{tab2} with other well known, exceptionally bright in the optical, galaxy-galaxy 
$\rm z \sim 3$ lenses: the Cosmic Eye \citep{smail}, MS 1512-cB58 \citep{yee}, the 8 o'clock \citep{allam}, and at lower 
redshift the Cosmic Horseshoe \citep{belo}. BG1429+1202 has similar brightness but is intrinsically very luminous in the 
rest-frame UV continuum and $\rm Ly \alpha$. It is also the only one $\rm Ly \alpha$ emitter \citep[the $\rm Ly \alpha$ 
line of the Cosmic Horseshoe galaxy shares many of the properties of the $\rm Ly \alpha$ emitters, but its $W_{0}^{\rm 
Ly \alpha}$ is below the threshold generally adopted to define it as $\rm Ly \alpha$ emitter;][]{quider2009}.
This puts BG1429+1202 in the small group of very bright $\rm z \sim 3$ galaxies that, due to the high magnification 
and its high intrinsic luminosity, their brightness provides the unique opportunity to obtain high S/N spectroscopy to 
study in detail its physical properties. 

\section{Conclusion} \label{sec: fin}

In this Letter, we report the discovery of a bright quadruply lensed LAE at $\rm z = 2.8224$. The very bright apparent 
magnitude results partially from gravitational lensing by a $\rm z=0.5531$ luminous red galaxy, which provides a 
magnification of $8.8 \pm 0.4$. After accounting for the lensing magnification, BG1429+1202 is also intrinsically very 
luminous in the rest-frame UV and Ly$\alpha$ emission by about 19 and 5 times the typical $L^{*}_{\rm UV}$ and 
$L^{*}_{\rm Ly\alpha}$ of LAEs at $z \sim 3$, respectively, showing low dust content and indications of massive recent 
star formation. Compared with the few well known strongly lensed galaxies,  it is the most luminous one in the Ly$\alpha$ 
line. This makes this source another good laboratory for further detailed studies of the physics of star formation 
and Ly$\alpha$ emission in galaxies during the cosmic epoch of star formation. The new method presented in 
\cite{shu_a,shu_b} and in this work opens a new window to the study of high redshift galaxies by the combination of 
massive spectroscopic surveys, large-area multi-band imaging, gravitational lensing and follow-up with 10 m telescopes 
like GTC.

\acknowledgments

We thank the anonymous referee for helpful comments.
We thank the IAC Director for the allocation of WHT and GTC discretionary time and the WHT and GTC staff for efficient 
support and fast completion of the two DDT programs. RMC acknowledges Fundaci\'on La Caixa for the financial support 
received in the form of a PhD contract. This work was funded in part by the project ESP2015-65597-C4-4-R of the Spanish 
Ministerio de Econom\'ia y Competitividad (MINECO).
SM and YS acknowledge the support by the Strategic Priority Research Program ``The Emergence of Cosmological 
Structures'' of the Chinese Academy of Sciences Grant No. XDB09000000 and by the National Natural Science Foundation 
of China (NSFC) under grant numbers 11333003, 11390372, and 11603032.
Funding for SDSS-III has been provided by the Alfred P. Sloan Foundation, the Participating Institutions, the National Science Foundation, and the U.S. Department of Energy Office of Science. The SDSS-III web site is http://www.sdss3.org/.

\end{document}